\documentclass[prl,twocolumn,showpacs,superscriptaddress]{revtex4}
\usepackage{graphicx}
\usepackage{amsmath}
\usepackage{amssymb}
\usepackage{color}

\begin{document}

\title{Absence of a boron isotope effect in the magnetic penetration
depth of MgB$_{2}$}
\author{D. Di Castro}
 \email[Email: ]{dicastro@physik.unizh.ch}
\affiliation{Physik-Institut der Universit\"at Z\"urich, CH-8057
Z\"urich, Switzerland}

\author{M. Angst}
\affiliation{Physik-Institut der Universit\"at Z\"urich, CH-8057
Z\"urich, Switzerland}

\author{D.~G. Eshchenko}
\author{R. Khasanov}
\affiliation{Physik-Institut der Universit\"at Z\"urich, CH-8057
Z\"urich, Switzerland}

\affiliation{Paul Scherrer Institute, CH-5232 Villigen PSI,
Switzerland}

\author{J. Roos}
\affiliation{Physik-Institut der Universit\"at Z\"urich, CH-8057
Z\"urich, Switzerland}

\author{I.~M. Savic}
\affiliation{Faculty of Physics, University of Belgrade, 11001
Belgrade, Yugoslavia}

\author{A. Shengelaya}
\affiliation{Physik-Institut der Universit\"at Z\"urich, CH-8057
Z\"urich, Switzerland}

\author{S. L. Bud'ko}
\author{P. C. Canfield}
\affiliation{Ames Laboratory and Department of Physics and Astronomy,
Iowa State University,Ames,IA 50011,USA}

\author{K. Conder}
\affiliation{Laboratory for Neutron Scattering, ETH Z\"{u}rich and
PSI Villigen, CH-5232 Villigen PSI}

\author{J. Karpinski}
\author{S.~M. Kazakov}
\affiliation{Solid State Physics Laboratory, ETH, CH-8093
Z\"urich, Switzerland}

\author{R. A. Ribeiro}
\affiliation{Ames Laboratory and Department of Physics and Astronomy,
Iowa State University,Ames,IA 50011,USA}

\author{H. Keller}
\affiliation{Physik-Institut der Universit\"at Z\"urich, CH-8057
Z\"urich, Switzerland}

%

%
\begin{abstract}
The magnetic penetration depth $\lambda(0)$ in polycrystalline
MgB$_{2}$ for different boron isotopes ($^{10}$B/$^{11}$B) was
investigated by transverse field muon spin rotation. No boron
isotope effect on the penetration depth $\lambda(0)$ was found
within experimental error: $ \Delta \lambda (0) / \lambda (0) =
-0.8(8) \%$, suggesting that MgB$_2$ is an adiabatic
superconductor. This is in contrast to the substantial oxygen
isotope effect on $\lambda(0)$ observed in cuprate
high-temperature superconductors.
\end{abstract}

\maketitle

%
%
Since the discovery of superconductivity with transition temperature
$T_{c} \approx
39\,{\mathrm{K}}$ in the
binary intermetallic compound MgB$_{2}$ \cite{Nagamatsu01}, a
large number of experimental and theoretical investigations were
performed in order to explain the mechanism and the origin of
its remarkably high  transition temperature. Experiments were done
revealing the important role played by the lattice excitations in
this material \cite{Bud'ko01,Hinks01,DiCastro02,Goncharov}.
In particular, the substitution of the $^{11}$B with $^{10}$B  has
been demonstrated to shift  $T_{c}$ to higher temperatures
\cite{Bud'ko01,Hinks01}, as expected for a phonon mediated pairing
mechanism.

However, MgB$_{2}$ differs from conventional  superconductors in
several important aspects, including, for instance, the unusually
high $T_{c}$ and the anomalous specific heat \cite{Bouquet01}.
Calculations \cite{Liu,Choi02} based on the Eliashberg formalism
support the experimental
results \cite{Bouquet01,Szabo01,Souma03}, revealing
MgB$_{2}$ to be a two-band superconductor with two
superconducting gaps of different size, the larger one originating
from a 2D $\sigma$-band and the smaller one from a 3D $\pi$-band.
The electronic $\sigma$-states are confined to the boron planes and
couple very strongly to the in-plane vibration of the boron atoms
($E_{2g}$ phonon mode). This strong pairing,  confined only to
parts of the Fermi surface, is the principal contribution
responsible for superconductivity and  mainly determines $T_{c}$.
The $\pi$-states on the remaining parts of the Fermi surface form
much weaker pairs. The double-gap structure explains most of the
unusual physical properties of MgB$_{2}$, such as the high
critical temperature, the total $T_{c}$ isotope-effect coefficient
($\alpha$ $\approx$ 0.32 \cite{Hinks01}), the temperature
dependent specific heat \cite{Bouquet01}, tunneling \cite{Szabo01}
and upper critical field anisotropy $H_{c2}^{\parallel
ab}$/$H_{c2}^{\parallel c}$ \cite{Angst02}.

An interesting point to be clarified concerns the
nature of the electron-lattice coupling. It was proposed
\cite{Alexandrov,Cappelluti02} that MgB$_{2}$ is a non-adiabatic
superconductor. Alexandrov \cite{Alexandrov} suggested that,
because of the large coupling strength of the electrons to the
E$_{2g}$ phonon mode, the many-electron system is unstable and
breaks down into a small polaron system, similar to the cuprate high
temperature superconductors (HTSC), where the charge carriers are
trapped by
 local lattice distortions. Cappelluti {\em et al.}
\cite{Cappelluti02} proposed that the small value of the Fermi energy
$E_{F}$ of the $\sigma$
 bands relative to the phonon energy $\omega_{ph}$ violates the
adiabatic
assumption ($\omega_{ph} \ll E_{F}$), opening up a non-adiabatic
channel that enhances $T_{c}$. These models are in contrast to the
conventional theory of superconductivity
 (Migdal adiabatic approximation),
 in which the density of states at the Fermi level,
the electron-phonon coupling constant, and the effective
supercarrier mass $m^{*}$ are all independent of the mass $M$ of
the lattice atoms. Both non-adiabatic models
 explicitly predict in MgB$_{2}$ a
boron isotope effect (BIE) on the charge carrier effective mass
$m^{*}$ \cite{Cappelluti02,Alexandrov}. Similar models
\cite{Alexandrov94,Grimaldi98} were
already used to explain the large oxygen isotope effect (OIE) on
$m^{*}$ observed in HTSC
\cite{Zhao95,Zhao97,Zhao98,Hofer00,Khasanov02,Khasanov03}
by measuring
 the OIE on the magnetic field penetration depth $\lambda$, a
physical quantity
directly related to the charge carrier effective mass.

 In this letter, a  muon spin rotation ($\mu$SR) study of the magnetic
 penetration depth $\lambda(0)$ in polycrystalline
MgB$_{2}$ for different boron isotopes ($^{10}$B/$^{11}$B) is
reported. $\mu$SR is a powerful microscopic tool to
measure the magnetic penetration depth $\lambda$ \cite{Pumpin}.
Indeed, in a polycrystalline type II superconductor with a perfect
vortex lattice (VL) the average magnetic penetration
depth $\lambda$ can be extracted from the muon-spin depolarization
rate $\sigma(T) \propto \lambda^{-2}(T)$ \cite{Pumpin}.
In our measurement, no BIE on  $\lambda(0)$ was observed within
experimental error [$\Delta\lambda(0)/\lambda(0)\!=\!-0.8(8)\%$],
in contrast to the substantial OIE observed in cuprate HTSC
\cite{Zhao95,Zhao97,Zhao98,Hofer00,Khasanov02,Khasanov03}.
Our results imply that polaronic or non-adiabatic effects in
MgB$_{2}$ are absent or
negligibly small.

The $\mu$SR experiments were performed on two polycrystalline
MgB$_{2}$ samples containing $^{11}$B (Mg$^{11}$B$_{2}$) and
$^{10}$B (Mg$^{10}$B$_{2}$). Full details of the sample synthesis
are given in Refs [\onlinecite{Bud'ko01,Ribeiro}]. In brief, the
two samples were synthesized using elemental Mg (99.9$\%$ pure in
lump form) and isotopically pure boron (99.95$\%$ chemical purity,
99.5$\%$ isotope purity, $<$100 mesh) combined in a sealed Ta tube
in a stoichiometric ratio. The Ta tube was then sealed in a quartz
ampoule, placed in a 950$^{\circ}$C box furnace for 24 hours, and
then removed and allowed to cool to room temperature.

\begin{figure}
\input{epsf}
\epsfysize 5.4cm \centerline{\epsfbox{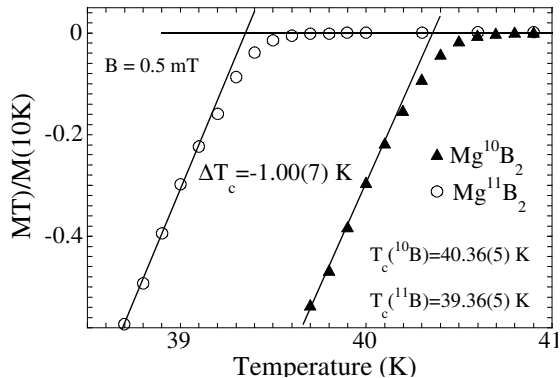}}
\caption[~]{Normalized field cooled ($0.5\,{\mathrm{mT}}$)
magnetization as a function of temperature for Mg$^{10}$B$_{2}$
and Mg$^{11}$B$_{2}$ samples.} \label{fig1}
\end{figure}

To examine the quality of the samples low field
($0.5\,{\mathrm{mT}}$, field-cooled) magnetization measurements
were performed using a commercial Superconducting Quantum
Interference Device. Figure \ref{fig1} shows the temperature
dependence of the magnetization for the Mg$^{11}$B$_{2}$ and
 Mg$^{10}$B$_{2}$ samples in
the vicinity of $T_{c}$. The high quality of the two samples is
revealed by the sharp transition and the high $T_{c}$
extracted from the intercept of the linear extrapolations (Fig.\
\ref{fig1}): $T_{c}(^{10}{\mathrm{B}}) = 40.36(5)\,{\mathrm{K}}$,
$T_{c}(^{11}{\mathrm{B}}) = 39.36(5)\,{\mathrm{K}}$. There is a
clear isotope shift of $\Delta T_{c} =
T_{c}(^{11}{\mathrm{B}})-T_{c}(^{10}{\mathrm{B}}) = -
1.00(7)\,{\mathrm{K}}$. The corresponding isotope effect
coefficient $\alpha_{B}=-dln(T_{c})/dln(M_{B}) = 0.29(2)$
(enrichment corrected) is in  good agreement with previous results
\cite{Bud'ko01,Hinks01}.

\begin{figure}
\input{epsf}
\epsfysize 5.4cm \centerline{\epsfbox{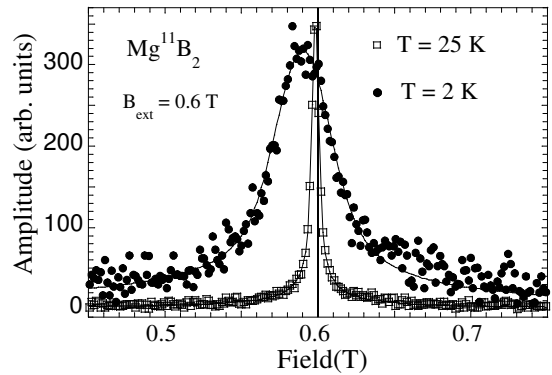}}
\caption[~]{Local magnetic field distribution, obtained from the
Fourier transform of the muon spin precession signal, for
Mg$^{11}$B$_{2}$ at $2$ ($\bullet$) and $25\,{\mathrm{K}}$
($\square$). Solid lines are gaussian fits to the experimental
data. The vertical solid line indicates the external field of
$0.6\,{\mathrm{T}}$.} \label{p(B)}
\end{figure}

The transverse-field  $\mu$SR  experiments were performed at the
Paul Scherrer Institute (PSI), Switzerland, using the $\pi$M3
$\mu$SR facility. The  samples used for the magnetization
measurements (see Fig.\ \ref{fig1}) were pressed in disk-shaped
pellets with $10\,{\mathrm{mm}}$ diameter and $3\,{\mathrm{mm}}$
thickness and cooled in an external magnetic field $B_{\rm{ext}}$
perpendicular to the muon spin polarization from well above
$T_{c}$ to temperatures lower than $T_{c}$. The measurements were
taken in a field of $B_{ext} = 0.6\,{\mathrm{T}}$ (the highest
available at PSI), high enough to avoid pinning induced distortion
of the VL \cite{note,Niedermayer02,Ohishi}.
As shown in Fig.\ \ref{p(B)} for Mg$^{11}$B$_{2}$ at two different
temperatures, the local magnetic field distribution can be very
well approximated by a single Gaussian, centered at a field lower
than the external one. This again indicates the high quality of
the samples and the absence of any normal conducting domains. From
the width of the Gaussian field distribution,
which is proportional to the muon spin depolarization
rate $\sigma$,
the penetration depth $\lambda$, that is the length scale of the
variation of the magnetic field,
can be extracted using the relation $\lambda^{-2}\propto\sigma$.

\begin{figure*}
\input{epsf}
\epsfysize 7.0cm \centerline{\epsfbox{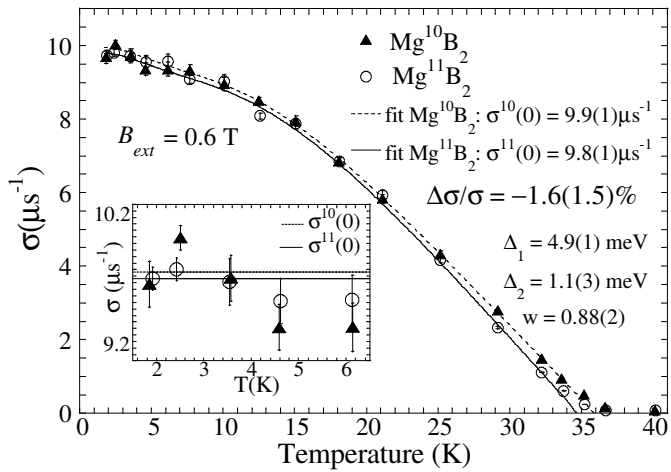}}
\caption[~]{Temperature dependence of 
$\sigma$ at $B_{\mathrm{ext}} = 0.6\,{\mathrm{T}}$ for the
two isotope samples Mg$^{10}$B$_{2}$ ($\blacktriangle$) and
Mg$^{11}$B$_{2}$ ($\circ$). The solid (Mg$^{10}$B$_{2}$) and
dotted (Mg$^{11}$B$_{2}$) lines are  fits using Eq.
(\ref{twogapmodel}). Inset: low-temperature region on
a larger scale. The dotted and solid horizontal lines represent
the weighed average values of $\sigma(0)$ for $T <
7.5\,{\mathrm{K}}$ for Mg$^{10}$B$_{2}$ and Mg$^{11}$B$_{2}$,
respectively.} \label{B11-B10_SvsT}
\end{figure*}

In Fig.\ \ref{B11-B10_SvsT}, the temperature dependence of
$\sigma$ for the Mg$^{11}$B$_{2}$ ($\circ$) and Mg$^{10}$B$_{2}$
($\blacktriangle$) samples is shown.
Below $T_{c}$, $\sigma$ for both  samples starts to increase  and
saturates at low temperatures $T \leq 6\,{\mathrm{K}}$, in
agreement with previous $\mu$SR measurements \cite{Niedermayer02}.
The data for the two samples close to $T_{c}$ show a clear isotope
shift of
 $\Delta T_{c} = - 1.2(2)\,{\mathrm{K}}$, in agreement
with $\Delta T_{c}$
 deduced from the low field magnetization measurements (Fig.
\ref{fig1}).
With  decreasing  temperature, the values of  $\sigma$ for the
 Mg$^{11}$B$_{2}$ sample are systematically lower than those for
 the Mg$^{10}$B$_{2}$ sample.
 However, at low temperature  they
 merge together, indicating that there is no substantial BIE on
 $\sigma(0)$.

In order to quantify
this observation, we performed fits to the experimental data.
It was suggested \cite{Niedermayer02,Ohishi} that for the
two-gap  superconductor MgB$_{2}$, the temperature dependence
of $\sigma$ can be written in
the form:
\begin{equation}
\sigma(T)=\sigma(0)- w \cdot \delta \sigma (\Delta_{1},T)-(1-w)
\cdot \delta \sigma (\Delta_{2},T) \label{twogapmodel}
\end{equation}
with $\delta\sigma(\Delta,T)$=$\frac{2\sigma (0)}{k_{B}
T}$$\int_{0}^{\infty}f(\varepsilon,T)$$\cdot$$[1-f(\varepsilon,T)]d\varepsilon$.

Here, $\Delta_{1}$ and $\Delta_{2}$ are the zero temperature large
and small gap, respectively, $w$ is the relative contribution of
the large gap to $\lambda^{-2}(0)$, and $f(\varepsilon,T)$ is the
Fermi distribution. For the temperature dependence of the gaps
we used the conventional BCS $\Delta(T)$.
In
order to improve the ratio of the number of data point $vs$ the
number of fit parameters, the two gaps and $w$ were considered as
common fitting parameters for the two isotope data.
As shown by the solid and dotted lines in Fig. \ref{B11-B10_SvsT},
the experimental data are  well described by Eq.\
(\ref{twogapmodel}). The fit yields:
$\sigma(0)^{^{11}{\mathrm{B}}} = 9.79(10) \mu{\mathrm{s}}^{-1}$,
$\sigma(0)^{^{10}{\mathrm{B}}} = 9.95(11) \mu{\mathrm{s}}^{-1}$,
$w$ = 0.88(2), $\Delta_{1}$ = 4.9(1) and $\Delta_{2}$ = 1.1(3).
All these values are in very good agreement with previous $\mu$SR
measurements performed by us on a natural boron MgB$_{2}$ sample
and by Ohishi {\em et al.} \cite{Ohishi}. It is interesting to
note that the high value of $w$ implies  that only a very small
contribution to $\sigma(0)$ originates from the $\pi$-band, in
accordance with the experimental finding that the superfluid
density in the $\pi$-band is strongly  suppressed by an external
magnetic field \cite{Bouquet01,Szabo01}.

The relative isotope shift of $\sigma(0)$ is:
$(\sigma(0)^{^{11}{\mathrm{B}}}\!-\!\sigma(0)^{^{10}{\mathrm{B}}})/\sigma(0)^{^{10}{\mathrm{B}}}
\equiv \Delta\sigma(0)/\sigma(0) =
\Delta\lambda^{-2}(0)/\lambda^{-2}(0) = -1.6(1.5)\%$,
corresponding to: 
\begin{equation}
\Delta\lambda(0)/\lambda(0) = -0.8(8)\%.
\label{lambda}
\end{equation}
For
comparison, we calculated the relative isotope shift using a
different and model independent procedure, taking the  weighed
average of the experimental points for $T < 7.5$ K (see inset of
Fig.\ \ref{B11-B10_SvsT}), where $\sigma(T)$ saturates. We
obtained $\Delta\sigma(0)/\sigma(0)$ =
$\Delta\lambda^{-2}(0)/\lambda^{-2}(0) = -0.5(8)\%$.
Both results are
compatible with zero BIE on
the penetration depth $\lambda(0)$.

Here, it is very important to recall that the two isotope samples
used in the experiment were made with the same starting Mg for both
the samples, and with  $^{10}$B
and $^{11}$B powders of the
same mesh size (distribution of grain sizes), and were synthesized
under
exactly the same conditions. Therefore, we can
exclude any influence on $\sigma$ due to different
grain size and to a difference in pinning or
vortex dynamical effects.

To check the reliability of our results, a second measurement on a
set (set B) of samples from different
source and preparation technique and with smaller Meissner
fraction, was performed in a field of $0.4\,{\mathrm{T}}$. The
results are very similar to the first set (set A) shown above:
$\Delta\lambda^{-2}(0)/\lambda^{-2}(0) = -1.5(1.7)\%$ as compared
to the above $-1.6(1.5)\%$. This shows that our result is
intrinsic for MgB$_2$ and holds for lower fields as well. A
summary of the results for both sets of isotope samples is given
in Table \ref{tablesummary}. Note that the values of
$\sigma(0)^{^{11}B}$ and $\sigma(0)^{^{10}B}$ for set B measured
in lower fields are larger than the corresponding values for set
A. This is consistent with previous work
\cite{Niedermayer02,Ohishi}.

\begin{table}[b]
\caption[~]{Summary of the BIE results for $\sigma(0)$ obtained
from the $\mu$SR measurements of two sets of isotope samples.}
\label{tablesummary}
\begin{ruledtabular}
\begin{tabular}{c|c|c|cc} 
%
&$\sigma(0)^{^{10}{\mathrm{B}}}$&
$\sigma(0)^{^{11}{\mathrm{B}}}$&\multicolumn{2}{c}{$\Delta\lambda^{-2}(0)/\lambda^{-2}(0)$}\\

&($\mu$s$^{-1}$)&($\mu$s$^{-1}$)&\\
\hline
Set
A&9.95(11)&9.79(10)&$-$0.016(15)\footnotemark\footnotetext[1]{from
fit using Eq.\ (\ref{twogapmodel})}&
$-$0.005(8)\footnotemark\footnotetext[2]{from low temperature average
(inset Fig.\ \ref{B11-B10_SvsT})}\\
Set B&12.91(17)&12.69(13)&$-$0.015(17)\footnotemark[1]&
$-$0.016(30)\footnotemark[2]\\
\end{tabular}
\end{ruledtabular}
\end{table}

Theoretically, the zero temperature penetration depth is
proportional to a density-of-states weighed average of a tensor
involving the Fermi velocities. Detailed calculations within
different formalisms have been carried out for MgB$_2$ (see Refs.\
\cite{Kogan02,Golubov02b}). For our purpose it is sufficient to
use the simpler London approach considering a free electron model
and linking $\lambda (0)$ to the superconducting charge carrier
density $n_{s}$ and effective mass $m^{*}$, only considering
different contributions from the $\sigma$ and the $\pi$ bands.
There is of course a direct connection between the Fermi
velocities and the effective mass (a band average), both of which
are not bare quantities, but in general renormalized, e.g., due to
coupling with the phonons. The London approach has the advantage
of facilitating the comparison with theoretical predictions
\cite{Alexandrov,Cappelluti02} and results obtained on cuprate
superconductors
\cite{Zhao95,Zhao97,Zhao98,Hofer00,Khasanov02,Khasanov03}, all of
which are formulated within this approach.

Unlike the cuprate superconductors with their extremely short
coherence lengths, MgB$_2$ cannot be considered as being in the
superclean limit and we need to consider a possible impact of
scattering.
In a moderately clean 
superconductor the penetration depth is related to the effective
mass $m^{*}$ by the following relation \cite{Tinkham}:
\begin{equation}
1/\lambda^2 = [\mu_{0} e^{2}/c^{2}] (n_{s}/m^{*}) [1/(1+\xi/\ell)],
\label{1/lambda}
\end{equation}
where $n_{s}$  and $m^{*}$ are  the superconducting charge carrier
density and effective mass, respectively, $\xi$ is the coherence
 length, and $\ell$ is the mean free path.
As already mentioned, the major contribution ($\sim 90 \%$) to $\lambda^{-2}$
in our
experimental conditions
 comes from the $\sigma$-band.
Therefore $n_{s}$, $m^{*}$, $\xi$, and  $\ell$ in Eq.(\ref{1/lambda})
have to be
considered as $\sigma$-band values.
It was estimated \cite{Sologubenko02,Bouquet02} that
 in the $\sigma$-band $(\xi/\ell)_{\sigma} \approx 1/8$, a
value which is close to the clean limit ($\xi/\ell \ll 1$). Therefore
Eq. (\ref{1/lambda}) may be approximated by
$1/\lambda^2 \approx [\mu_{0} e^{2}/c^{2}] (n_{s}/m^{*})$.
A shift in $1/\lambda^2$ due to the
isotope substitution is then given
by
\begin{equation}
\frac{\Delta\lambda^{-2}(0)}{\lambda^{-2}(0)} =\frac{\Delta
n_{s}}{n_{s}}-\frac{\Delta m^{\ast}}{m^{\ast}}.
\label{IE}
\end{equation}
The contribution from the supercarrier density $n_{s}$ is
negligible, as  was already experimentally demonstrated in the
case of HTSC \cite{Zhao97,Zhao98,Hofer00,Khasanov03}.
Specifically, for MgB$_{2}$, it can be argued that: i) by
changing the isotope only the mass of the nuclei is changed and
not the charge carrier density $n$. Furthermore, MgB$_{2}$ is a
stoichiometric compound; ii)  x-ray diffraction
measurements, performed on the samples used for  the $\mu$SR
experiments, showed  no substantial difference between the lattice
parameters of Mg$^{11}$B$_{2}$ and Mg$^{10}$B$_{2}$.
This implies that the band structure is not  appreciably
modified by the isotope substitution.
Therefore,
assuming $\Delta n_{s}/n_{s} \approx 0$ in Eq.(\ref{IE}) and
neglecting the small $\pi$-band contribution, we can estimate the
boron isotope
effect on the $\sigma$-band effective mass $m^{*}_{\sigma}$:
\begin{equation}
\Delta m^{\ast}_{\sigma}/m^{\ast}_{\sigma} \approx
-\Delta\lambda^{-2}(0)/\lambda^{-2}(0) =1.6(1.5)\%.
\label{DeltaM}
\end{equation}
Here we  have used the value of the relative shift on
$\lambda^{-2}(0)$ obtained from the fit to  Eq.
(\ref{twogapmodel}).
There is no BIE on the $\sigma$-band effective mass within
experimental error.

Our result then suggests that
non-adiabatic or polaronic effects in MgB$_{2}$  are absent or
negligibly small, and establishes an upper
limit [Eq. (\ref{lambda}) and
          Eq. (\ref{DeltaM})]
to any theoretical prediction of such effects \cite{Alexandrov,Cappelluti02}.
This conclusion is in contrast to cuprate superconductors, where a substantial
oxygen
isotope effect on $m^{\ast}$ was observed
\cite{Zhao95,Zhao97,Zhao98,Hofer00,Khasanov02,Khasanov03}.
Recent magnetization measurements on
MgB$_{2}$ under pressure \cite{DiCastro03} show no pressure effect
on the magnetic penetration depth $\lambda$ at low temperature,
further supporting the main conclusion of the present work.

In summary, $\mu$SR experiments on polycrystalline
Mg$^{10}$B$_{2}$ and Mg$^{11}$B$_{2}$ samples revealed no
substantial boron isotope effect on the
 magnetic
penetration depth at $T = 0\,{\mathrm{K}}$. From this finding we
conclude that there is
 no substantial BIE  on the  effective mass $m^{\ast}_{\sigma}$ of the
 charge carriers in the $\sigma$ band. This result suggests that
MgB$_{2}$ is a conventional phonon mediated superconductor without
non-adiabatic or polaronic effects, in contrast to
cuprate superconductors.

This work was partly performed at
the Swiss Muon Source (S$\mu$S) at the Paul Scherrer Institute
(Villigen, Switzerland). We thank D. Herlach and A. Amato for
technical assistance during the
$\mu$SR experiments at the Paul Scherrer Institute and T. Schneider
for useful discussions. This work was
supported by the Swiss National Science Foundation and by the NCCR
Program MaNEP sponsored by the Swiss National Science Foundation.
Ames Lab is operated for the U.S.
Department of Energy by Iowa State University under Contract No.
W-7405-Eng-82.  The work at the Ames Lab was supported by the
Director of Energy Research,
Office of Basic Energy Sciences.

\newcommand{\noopsort}[1]{} \newcommand{\printfirst}[2]{#1}
  \newcommand{\singleletter}[1]{#1} \newcommand{\switchargs}[2]{#2#1}


\begin{thebibliography}{10}
\bibitem{note} Random flux pinning near the lower
critical field H$_{c1}$ may induce distortion of the VL, giving
rise to a field dependent $\sigma$ which affects
 the value of $\lambda$.
\end{thebibliography}
\end{document}